%
%
%
\magnification=\magstephalf
\hyphenation{brem-sstrahlung}
\hsize=6.5truein 
\vsize=9.15truein
\hoffset=-0.1truein
\voffset=-.15truein 

\baselineskip=24truept plus 0truept minus 0truept

\parindent=3em 
\overfullrule=0pt
\headline={\ifodd\pageno\rightheadline\else\rightheadline\fi}
\def\numbers{\def\rightheadline{\hfil\tenrm\folio\hfil}} 
\def\nonumbers{\def\rightheadline{\hfil}} 
\nonumbers 
\nopagenumbers 
\tolerance=800 \hbadness=5000
\newdimen\listindent\listindent=2em 

\def\refs{\leftskip=2em\parindent=-2em}
\def\sig{\sigma_{_{\rm T}}}
\def\Max{{\rm Max}}

\def\msun{M_\odot}
\def\gapprox{\lower.4ex\hbox{$\;\buildrel >\over{\scriptstyle\sim}\;$}}
\def\lapprox{\lower.4ex\hbox{$\;\buildrel <\over{\scriptstyle\sim}\;$}}

\def\ab{\alpha_{\gamma\gamma}}
\def\ta{\tau_{\gamma\gamma}}
\def\si{\sigma_{\gamma\gamma}}
\def\in{I_\epsilon}
\def\epsmin{{\epsilon_{\rm min}(\theta)}}

\def\t0{{\theta_0}}
\def\zg{z_{\gamma\gamma}}
\def\zcool{{z_{\rm cool}}}
\def\thetamin{{\theta_{\rm min}(z)}}
\def\thetamax{{\theta_{\rm max}(z)}}
\def\rms{R_{\rm ms}}
\def\r0{R_0}
\def\rg{R_g}
\def\redshift{\zeta}
\def\redshift{{\cal Z}}
\def\tg{t_{\gamma\gamma}}

\vskip1.5truein
\centerline{\bf IMPLICATIONS OF GAMMA-RAY TRANSPARENCY}
\centerline{\bf CONSTRAINTS IN BLAZARS: MINIMUM DISTANCES}
\centerline{\bf AND GAMMA-RAY COLLIMATION}
\bigskip
\bigskip
\centerline{Peter A. Becker and Menas Kafatos}
\bigskip
\centerline{Center for Earth Observing and Space Research,}
\centerline{Institute for Computational Sciences and Informatics,}
\centerline{and Department of Physics and Astronomy,}
\centerline{George Mason University, Fairfax, VA 22030-4444}
\bigskip
\bigskip
\bigskip
\bigskip
\vskip2.0truein
\centerline{(This paper appeared in the November 1, 1995 edition of the
{\it Astrophysical Journal})}
\vskip 3.0truecm
\vfil
\eject

\numbers
\pageno=2
\bigskip
\centerline{ABSTRACT}

We develop a general expression for the $\gamma-\gamma$ absorption
coefficient, $\ab$, for $\gamma$-rays propagating in an arbitrary
direction at an arbitrary point in space above an X-ray emitting
accretion disk. The X-ray intensity is assumed to vary as a power
law in energy and radius between the outer disk radius, $\r0$,
and the inner radius, $\rms$, which is the radius of marginal
stability for a Schwarzschild black hole.
We use our result for $\ab$ to calculate the $\gamma-\gamma$ optical
depth, $\ta$, for $\gamma$-rays created at height $z$ and propagating
at angle $\Phi$ relative to the disk axis, and we show that for
$\Phi=0$ and $z\gapprox \r0$, $\ta \propto E^\alpha\,z^{-2\alpha-3}$,
where $\alpha$ is the X-ray spectral index and $E$ is the $\gamma$-ray
energy. As an application, we use our formalism to compute the
minimum distance between the central black hole and the
site of production of the $\gamma$-rays detected by EGRET
during the June 1991 flare of 3C~279. In order to obtain an upper
limit, we assume that all of the X-rays observed contemporaneously
by {\it Ginga} were emitted by the disk. Our results suggest that
the observed $\gamma$-rays may have originated within $\lapprox
45\,GM/c^2$ from a black hole of mass $\gapprox 10^9\,\msun$, perhaps
in active plasma located above the central funnel of the accretion disk.
This raises the possibility of establishing a direct connection between
the production of the observed $\gamma$-rays and the accretion of material
onto the black hole. We also consider the variation of the optical depth
as a function of the angle of propagation $\Phi$. Our results indicate
that the ``focusing'' of the $\gamma$-rays along the disk axis due to
pair production is strong enough to explain the observed degree of
alignment in blazar sources. If the $\gamma$-rays are produced
isotropically in $\gamma$-ray blazars, then these objects should
appear as bright MeV sources when viewed along off-axis lines of sight.

\vfil
\eject

\bigskip
\centerline{\bf 1. INTRODUCTION}
\bigskip

Over the past several years, the EGRET instrument on board the
{\it Compton Gamma Ray Observatory} (CGRO) has detected high-energy
$\gamma$-rays from 44 active galactic nuclei (AGNs), and several
of these objects have also been detected at lower energies by
COMPTEL and OSSE (Fichtel et al. 1994; Hartman et al. 1992;
Hermsen et al. 1993; McNaron-Brown et al. 1994; von Montigny
et al. 1995). One of the most dramatic
events seen was the intense $\gamma$-ray flare of 3C~279 observed
in June 1991, which has called into question the viability of a
number of theoretical models for $\gamma$-ray emission in AGNs.
Most of the energy emitted during the flare appeared at energies
exceeding $\sim 100\,$MeV, with an implied (isotropic) luminosity
of $\sim 10^{48}\,{\rm ergs\,s}^{-1}$.
However, the
actual luminosity may have been several orders
of magnitude lower if the emission was not isotropic, but rather
``focused'' onto a small region of the sky due to some alignment
mechanism such as relativistic beaming or geometrical collimation.
In particular, the hypothesis of relativistic motion is consistent
with observations of apparent superluminal motion, flat radio spectra,
rapid optical variability, and high polarization in blazars.


In addition to beaming and geometrical collimation, $\gamma-\gamma$
pair production in the X-ray field of the accretion disk could in
principle lead to preferential escape of the $\gamma$-rays along the
symmetry axis of the disk, due to the strong angular dependence of
the pair-production cross section. The phenomenon of $\gamma-\gamma$
``focusing'' is related to the more general issue of $\gamma-\gamma$
transparency, which sets a minimum distance between the central black
hole and the site of $\gamma$-ray production, $\zg$. One usually finds
that $\zg$ greatly exceeds the cooling length,
$$
\zcool = c\,E\left({dE\over dt}\biggm|_{\rm synch}
+ {dE\over dt}\biggm|_{\rm IC}\right)^{-1}\,, \eqno(1.1)
$$
for relativistic electrons of energy $E$ traveling at essentially
the speed of light $c$ and suffering inverse-Compton and synchrotron
losses in the intense magnetic and radiation fields close to the
center of the active nucleus (Dermer \& Schlickeiser 1993). An
efficient pair creation and/or reacceleration mechanism is therefore
required in order to explain the presence of relativistic electrons
sufficiently far from the black hole to produce observable $\gamma$-rays.
The nature of this mechanism is unclear, and a variety of
possibilities have been suggested, including shocks propagating along
a jet (Blandford 1993), the decay of charged pions resulting from
hadronic collisions (Bednarek 1993), and the acceleration of electrons
and positrons via wave-particle interactions in the funnel of a thick
accretion disk (Becker, Kafatos, \& Maisack 1994). Constraints on the
mechanism are provided by estimates of the distance over which it must
operate, $\zg-\zcool$. Clearly, the closer one can place the $\gamma$-ray
production site to the black hole while maintaining $\gamma$-ray
transparency, the easier it will be to power the observed emission
using accretion. It is therefore interesting to calculate the minimum
distance satisfying the $\gamma$-ray transparency constraint in the
context of a specific X-ray emission scenario, and that is the focus
of this paper.

Gould \& Schr\'eder (1967) calculated the absorption coefficient
($\ab\propto{\rm cm}^{-1}$) describing the process $\gamma + \gamma'
\to e^+ + e^-$ for high-energy photons traversing a soft photon gas,
and their formalism has been used subsequently by Blandford (1993),
Dermer \& Schlickeiser (1994), and Bednarek (1993) to estimate the
absorption probability for $\gamma$-rays traversing the X-ray field
of an AGN. The detailed results depend strongly on the energy and
angular distribution of the soft radiation, and are therefore sensitive
to the presence of scattered radiation, which is more isotropic than the
primary radiation emitted by the disk. Blandford (1993) and Dermer \&
Schlickeiser (1994) assumed that the soft photon distribution is isotropic,
which is reasonable if the X-ray distribution is dominated by the
scattered component.
However, the relative importance of
the scattered photons compared to the primary (accretion disk)
photons is determined by the electron scattering optical depth
above the disk,
which is not known with any certainty. Bednarek
(1993) adopted the opposite viewpoint in his calculation of the
optical depth to pair production ($\ta$), neglecting electron
scattering and instead treating the full anisotropy of the soft
radiation field produced by the accretion disk. He assumed a
standard disk radiating as a blackbody with local temperature
$T(R)\propto R^{-3/4}$. The resulting X-ray intensity (integrated
over the face of the disk) has the characteristic power-law shape
$I_\nu\propto\nu^{1/3}$ (e.g., Pringle 1981). This is inconsistent
with most AGN (and blazar) X-ray spectra (with $I_\nu\propto\nu^{-1/2}$
typically), which are usually better fit by hot, two-temperature
accretion disk spectra (e.g., Shapiro, Lightman, \& Eardley 1976,
hereafter SLE) in the energy range relevant for pair production
($\epsilon > 261\,$eV for $\gamma$-ray energy $E=1\,$GeV;
see eq.~[3.4]).


In $\gamma$-ray loud blazars such as 3C~279, variability arguments
suggest that a significant fraction of the observed X-rays are probably
produced in a relativistic beam moving toward the observer (Makino
et al. 1989).
Since these X-rays are coaligned with the observed high-energy
$\gamma$-rays, they do not contribute substantially to the
$\gamma-\gamma$ opacity, and therefore they do not strongly constrain
the minimum distance between the black hole and the site of $\gamma$-ray
production. However, as we demonstrate below, an estimate of this
distance can be obtained by considering the attenuation of the
$\gamma$-rays due to collisions with X-rays emitted by an underlying
accretion disk. In order to tie the distance estimate
we obtain to the multifrequency observations of particular objects such
as 3C~279 without incurring additional theoretical complexity, in this
paper we make the further simplifying assumption that {\it all} of the
contemporaneously observed X-rays are emitted by the accretion disk.
By {\it overestimating} the X-ray flux emitted by the disk, we obtain
an ``upper limit'' on the minimum distance between the black hole and
the point of creation of the observed $\gamma$-rays, and we expect that
a proper accounting for the component of the X-ray flux generated in the
relativistic beam would further reduce this distance.
In any event, the upper limit we obtain
is small enough to allow for the possibility of establishing a direct
connection between the production of the observed high-energy
$\gamma$-rays and the accretion of matter onto the black hole.

\bigskip
\centerline{\bf 2. X-RAY INTENSITY OF THE DISK}
\bigskip

Electrons in the hot, inner region of a
two-temperature accretion disk produce power-law inverse-Compton
X-ray emission by upscattering UV radiation produced in the cool
(single-temperature) surrounding region. The temperature of the electrons
in the hot region is maintained at a nearly uniform value in the range
$T_e \sim 10^9-10^{10}\,$K by the ``Compton thermostat'' mechanism
(SLE; Eilek \& Kafatos 1983, hereafter EK), and the emitted X-ray spectrum
has the characteristic power-law shape
$$
I_\epsilon\propto\epsilon^{-\alpha}\,\,,\ \ \ \ \ \ \ 
\alpha=-{3\over 2} + \sqrt{{9\over 4} + {4\over y}}\,\,, \eqno(2.1)
$$
in the frame of the source, where $y\equiv(4 k T_e/m_e c^2)
\Max(\tau_{\rm es},\tau^2_{\rm es})\sim 1$ is the Compton
$y$-parameter, $m_e$ is the electron mass, and $\tau_{\rm es}$
is the electron-scattering half-thickness of the disk.
The power-law extends up to photon energies $\epsilon\sim kT_e$, above
which the spectrum declines exponentially. Since $\alpha$ and $T_e$ are
essentially constant in the hot inner region of the disk, it follows
that the radial variation of the X-ray intensity follows that of the
total flux, given by
$$
F(R)={3\over 8\pi}{GM\dot M\over R^3}\left[1-\left(6GM\over c^2 R
\right)^{1/2}\right]\,, \eqno(2.2)
$$
for a quasi-Keplerian accretion disk around a Schwarzschild black
hole of mass $M$, where $\dot M$ is the accretion rate (Shakura \&
Sunyaev 1973). Note that the total flux has a nearly power-law
dependence on radius $F(R)\propto R^{-3}$, with a sharp cutoff
at the radius of marginal stability for circular orbits, $\rms=
6\,\rg$, where $\rg\equiv GM/c^2$.

Based upon equations~(2.1) and (2.2), we conclude that the X-ray
intensity in the frame of the disk can be represented
approximately by
$$
\in(\epsilon,R) = I_0 \left(\epsilon\over m_e c^2\right)^{-\alpha}
\left(R\over\rg\right)^{-\omega}\,\,\,\,\,\,\,\,\,\,
(\propto{\rm ergs\,s^{-1}\,cm^{-2}\,ster^{-1}\,erg^{-1}})\,,
\eqno(2.3)
$$
where $\alpha$ is given by equation~(2.1), $\r0$ is the outer
radius of the two-temperature region, and $\rms \le R \le \r0$.
According to equation~(2.2),
$\omega=3$ for a two-temperature disk, but we shall leave $\omega$
as a free parameter so that we can make comparisons with uniformly-bright
disks, with $\omega=0$. Since it is unclear to what degree electron
scattering isotropizes the X-rays above the disk, we follow
Bednarek's approach, and neglect the effect of electron scattering
on the angular distribution of the X-rays. Our neglect of a scattered
X-ray component is in some sense self-consistent, since it turns out
that most of the $\gamma-\gamma$ attenuation occurs very close to the
disk, where the X-ray distribution is thought to be dominated by
unscattered accretion disk photons (Dermer \& Schlickeiser 1994).
However, the scattered radiation may attenuate
the $\gamma$-rays over much larger length scales.

Although the inner region of the disk terminates at the radius of
marginal stability, the
outer radius $\r0$ of the two-temperature region is more difficult
to quantify because of its dependence on the value (and possible variation)
of the viscosity parameter $\alpha$, but one typically finds
$30\,\rg\lapprox \r0\lapprox 100\,\rg$ (SLE; EK). Most of the photons
emitted in the ``cool'' region beyond radius $\r0$ are too soft to produce
pairs via collisions with $\gamma$-rays of energy $E \sim 1\,$GeV, and
therefore the effect of these photons on the $\gamma-\gamma$ absorption
coefficient is negligible.

In order to link our calculation of the $\gamma-\gamma$ absorption
coefficient with the X-ray and $\gamma$-ray data, we must express
the disk-frame intensity parameter $I_0$ in equation~(2.3) in terms of
the parameter $F'_0$ characterizing the observed (Doppler-shifted) X-ray
flux,
$$
F'_\epsilon = F'_0 \left(\epsilon'\over m_e c^2\right)^{-\alpha}
\,\,\,\,\,\,\,\,\,\,
(\propto{\rm s^{-1}\,cm^{-2}})\,,
\eqno(2.4)
$$
which is related to the observed intensity
$$
\in'(\epsilon',R) = I'_0 \left(\epsilon'\over m_e c^2\right)^{-\alpha}
\left(R\over\rg\right)^{-\omega}\,,
\eqno(2.5)
$$
by
$$
F'_\epsilon = \int \in' \cos\theta'\,d\Omega'\,, \eqno(2.6)
$$
where primes denote quantities measured in the frame of the earth.

Since we are mainly interested in blazar sources, where we are
presumably viewing the accretion disks face-on, we can express
$F'_0$ in terms of $I'_0$ and the luminosity distance to the source
$D$ by substituting for the intensity in equation~(2.6) using
equation~(2.5) and integrating over the face of the disk. Making
the substitutions $d\Omega'=2\pi\sin\theta'\,d\theta'$ and
$\tan\theta'=R/D$ yields
$$
F'_0 = 2 \pi D^2 I'_0 \int_{\rms}^{\r0}
\left(R\over\rg\right)^{-\omega}
{R\,dR \over (R^2 + D^2)^2}\,, \eqno(2.7)
$$
or, in the limit $D \gg \r0$,
$$
F'_0 = {2 \pi I'_0\over 2-\omega} \left(D\over\rg\right)^{-2}
\left[\left(\r0\over\rg\right)^{2-\omega}
-\left(\rms\over\rg\right)^{2-\omega}\right] \eqno(2.8)
$$
for $\omega\ne 2$, where $I'_0$ is related to $I_0$ via the Lorentz
transformation of the intensity, which gives for a source located at
redshift $\redshift$ (e.g., Begelman, Blandford, \& Rees 1984)
$$
I'_0 = (1+\redshift)^{-3-\alpha}\,I_0\,.
\eqno(2.9)
$$
In \S~3 we obtain a
general formula for the $\gamma-\gamma$ absorption coefficient for
$\gamma$-rays propagating above an accretion disk radiating with
the X-ray intensity distribution given by equation~(2.3).

\bigskip
\centerline{\bf 3. PHOTON-PHOTON PAIR PRODUCTION}
\bigskip

The exact cross section for $\gamma-\gamma$ pair production is given
by the well-known result (e.g., Jauch \& Rohrlich 1955),
$$
\si(\beta) = {3\over 16}\sig (1-\beta^2)\left[
(3-\beta^4)\ln\left(1+\beta\over 1-\beta\right) - 2\beta
(2-\beta^2)\right]\,, \eqno(3.1)
$$
where $\beta$ is the velocity of the positron (or electron)
in the center-of-momentum (CM) frame in units of $c$, and
$\sig$ is the Thomson cross section. The Lorentz
factor of the particles in the CM frame, $\Gamma$, is related to the
energy of the soft photon $\epsilon$, the energy of the
$\gamma$-ray $E$, and the interaction angle between the photon
trajectories $\theta$ by
$$
\Gamma^2 = {1\over 1-\beta^2}
= {\epsilon E (1-\cos\theta) \over 2 m_e^2 c^4}\,,
\eqno(3.2)
$$
where $\epsilon$, $E$, and $\theta$
are all measured in the same (arbitrary) frame. The cross section
for the process, $\si$, vanishes as $\Gamma \to 1$. Hence for fixed
$\epsilon$ and $E$, the absorption probability decreases as $\theta
\to 0$, and consequently the optical depth to pair production ($\ta$)
drops sharply as the photons become co-aligned. We therefore anticipate
that the value of $\ta$ will depend sensitively on the angular distribution
of the soft radiation relative to the direction of propagation of the
$\gamma$-ray.

\medskip
\centerline{ 3.1. \it $\gamma-\gamma$ Absorption Coefficient}
\medskip

Working in terms of the specific intensity $\in$, we can write the
differential contribution to the absorption coefficient $\ab$ due to
soft photons propagating in energy range $d\epsilon$ and in
directional range $d\Omega$ as
$$
d\ab = {\in\over c \epsilon}\,\si(\beta)\,(1-\cos\theta)\,
d\epsilon\,d\Omega\,. \eqno(3.3)
$$
We assume that the X-ray intensity of the disk is given by equation~(2.3),
and that any change in the X-ray spectral index $\alpha$ occurs below
the minimum soft-photon energy required to produce a pair with $\Gamma=1$,
$$
\epsmin = {2\,m_e^2 c^4 \over E\,(1-\cos\theta)}\,. \eqno(3.4)
$$
Taking $E=1\,$GeV, we find that $\epsilon_{\rm min}=0.522\,$keV for
a right-angle collision ($\cos\theta=0$), and $\epsilon_{\rm min}
=0.261\,$keV for a head-on collision ($\cos\theta=-1$), which is the
minimum possible X-ray energy resulting in pair production. The X-ray
spectrum given by equation~(2.3) would conflict with the measured
EGRET spectrum if extended to arbitrarily high energies, and
consequently a break must exist in the
energy range $0.1-1\,$GeV. However, for our purposes, the X-ray power-law
can formally be extended to infinite energy without producing numerical
divergence because $\si\propto\epsilon^{-1}\ln\epsilon$ as $\epsilon\to
\infty$ for fixed $E$ and $\theta$.

Using equation~(2.3) for the intensity, integration of equation~(3.3)
over $\epsilon$ and $\Omega$ yields for the absorption coefficient
$$
\ab = {I_0\over m_e c^3} \int \int_\epsmin^\infty
\left(\epsilon\over m_e c^2\right)^{-1-\alpha}
\left(R\over\rg\right)^{-\omega}
\si(\beta)\,(1-\cos\theta)\,d\epsilon
\,d\Omega\,, \eqno (3.5)
$$
where $\theta$ is the angle between $d\Omega$ and the direction
of propagation of the $\gamma$-ray, and the domain for the $\Omega$
integration corresponds to the range of solid angles subtended by
the disk. Equation~(3.5) can be broken into two separate
integrals by transforming from $\epsilon$ to $\beta$ using equation~(3.2).
The result is
$$
\ab = {I_0\over c}\left(E\over 2 m_e c^2\right)^\alpha
\sig \Psi(\alpha)
\,\int (1-\cos\theta)^{\alpha+1}
\left(R\over\rg\right)^{-\omega}\,d\Omega\,, \eqno (3.6)
$$
where the integration over $\beta$ is expressed by the function
$$
\Psi(\alpha)\equiv \sig^{-1}\int_0^1 2\beta\,(1-\beta^2)^{\alpha-1}\,
\si(\beta)\,d\beta\,, \eqno (3.7)
$$
plotted in Figure~1.

Equation~(3.6) can be used to calculate the absorption coefficient
for a $\gamma$-ray propagating in an {\it arbitrary} direction at
an {\it arbitrary} point in space. However, it is sufficient for
our purposes to consider $\gamma$-rays propagating in a plane that
includes the symmetry axis of the disk. Referring to the geometry
indicated schematically in Figure~2, we assume that the $\gamma$-ray
propagates in the $(x,z)$ plane, where the $z$-axis is the symmetry
axis of the disk, and we set the azimuthal angle $\varphi=0$ along
the $x$-axis. In general, we wish to relate $d\Omega$ to the differential
area in the disk $dA=R\,dR\,d\varphi$ using
$$
d\Omega = {\cos\eta~R\,dR\,d\varphi\over\ell^2}\,, \eqno(3.8)
$$
where $\eta$ is the angle between the X-ray trajectory and the
normal to the disk surface, and $\ell$ is the distance between the
photon-photon interaction point and the X-ray emission point in the
disk. Representing the location of the $\gamma$-ray using polar angle
$\delta$ and distance from the origin $\rho$, and assuming that the
$\gamma$-ray trajectory makes an angle $\Phi$ with the $z$-axis (see
Fig.~2), our expression for the absorption coefficient becomes
$$
\ab(E,\rho,\delta,\Phi) = {I_0\over c}\left(E\over 2 m_e c^2\right)^\alpha
\sig \Psi(\alpha)
\,\int_{\rms}^{\r0} \int_0^{2 \pi} (1-\cos\theta)^{\alpha+1}
\left(R\over\rg\right)^{-\omega}\,
{\cos\eta\,d\varphi\,R dR\over\ell^2}\,, \eqno (3.9)
$$
where
$$
\ell^2 = \rho^2 + R^2 - 2\,\rho\,R \sin\delta\cos\varphi\,, 
\ \ \ \ \ \ \ 
\cos\theta = {\rho\,\cos(\delta-\Phi)-R\,\cos\varphi\sin\Phi\over\ell}\,, 
\ \ \ \ \ \ \ 
\cos\eta = {\rho\,\cos\delta\over\ell}\,. \eqno(3.10)
$$
Note that $\ab\propto E^\alpha$, which verifies that softer X-ray
spectra yield larger optical depths. Equations~(3.9) and (3.10)
comprise our fundamental result for the $\gamma-\gamma$ absorption
coefficient for a $\gamma$-ray of energy $E$ propagating in the
power-law X-ray field of an accretion disk.

\medskip
\centerline{ 3.2. \it Absorption Coefficient for Propagation Along
the Disk Axis}
\medskip

One of the most interesting cases to consider is that of the attenuation
of $\gamma$-rays propagating along the symmetry axis of the disk, since
this scenario is of direct relevance for the interpretation of the
$\gamma$-rays detected by EGRET from blazar sources. In this case
equation~(3.9) is simplified substantially due to the cylindrical
symmetry, and the integration over $\varphi$ is trivial. We can obtain
the absorption coefficient for a $\gamma$-ray of energy $E$ propagating
outward along the symmetry axis at height $z$ by setting $\rho=z$ and
$\delta=\Phi=0$ in equation~(3.9) and transforming from $R$ to $\theta$
using $\tan\theta=R/z$, yielding
$$
\ab(E,z) = {2 \pi\,I_0\over c}\left(E\over 2 m_e c^2\right)^\alpha
\sig \Psi(\alpha)
\left(z \over\rg\right)^{-\omega}
\int_\thetamin^\thetamax (1-\cos\theta)^{\alpha+1}
\,\tan^{-\omega}\theta\,\sin\theta\,d\theta\,, \eqno (3.11)
$$
where $\thetamax$ [$\thetamin$] is the maximum (minimum) value of
the interaction angle $\theta$ at height $z$. Neglecting
gravitational curvature of the photon trajectories, we have
$$
\cos\thetamax = {z \over \sqrt{z^2 + \r0^2}}\,, \ \ \ \ \ \ \ \ 
\cos\thetamin = {z \over \sqrt{z^2 + R_{\rm ms}^2}}\,. \eqno (3.12)
$$
By writing equation~(3.11), we have implicitly assumed that $z \gg h$,
where $h$ is the half-thickness of the disk. This condition is trivially
satisfied for a thin disk, but the disk may also be considered to be thick,
so long as we confine our attention to points sufficiently far above the
surface.

\bigskip
\centerline{\bf 4. CALCULATION OF THE OPTICAL DEPTH}
\bigskip

We can use equations~(3.9) and (3.10) to calculate the $\gamma-\gamma$
optical depth $\ta$ for a $\gamma$-ray of energy $E$ created at radius
$R$ and height $z$ and propagating at angle $\Phi$ with respect to
the $z$-axis by writing
$$
\ta(E,R,z,\Phi) = \int_0^\infty \ab(E,\rho,\delta,\Phi)\,d\lambda\,,
\eqno(4.1)
$$
where $\lambda$ is the distance traversed by the $\gamma$-ray since
its creation, and $\rho$ and $\delta$ are given by
$$
\rho^2 = R^2 + z^2 + \lambda^2 + 2\,\lambda\,
(R\,\sin\Phi + z\,\cos\Phi)\,, \ \ \ \ \ \ \ \ 
\sin\delta = {R + \lambda\,\sin\Phi\over\rho}\,, \ \ \ \ \ \ \ \ \ \ \ 
\cos\delta = {z + \lambda\,\cos\Phi\over\rho}\,. \eqno(4.2)
$$
In keeping with equation~(3.9), we have assumed that the plane of
propagation of the $\gamma$-ray includes the $z$-axis.

\medskip
\centerline{4.1. \it Propagation Along the Disk Axis}
\medskip

We can obtain some further simplification for $\gamma$-rays
propagating along the $z$-axis, since then $\delta=\Phi=R=0$,
and $\rho=z + \lambda$. The optical depth to pair
production for a $\gamma$-ray of energy $E$ created at height
$z$ above the center of the disk and propagating outward along
the $z$-axis is then given by
$$
\ta(E,z) = \int_z^\infty \ab(E,z_*)\,dz_*\,, \eqno(4.3)
$$
where $\ab$ is given by equation~(3.11). Unfortunately, one
must resort to multidimensional numerical integration to compute
$\ta$ even in this highly symmetrical case. However, it is possible
to obtain a closed-form expression for $\ta$ if $z \gapprox \r0$.
Approximating the angular integral in equation~(3.11) yields for
the absorption coefficient
$$
\ab(E,z) \cong A\left(z\over\rg\right)^{-2\alpha-4}
\left(E\over 4 m_e c^2\right)^\alpha\,, \eqno (4.4)
$$
where
$$
A \equiv {\pi\,I_0\,\sig\,\Psi(\alpha)
\over (2\alpha + 4 - \omega)\,c}
\left[\left(\r0\over\rg\right)^{2\alpha+4-\omega}
- \left(\rms\over\rg\right)^{2\alpha+4-\omega}\right]\,. \eqno (4.5)
$$
Substituting equation~(4.4) into equation~(4.3) then yields the
approximate far-field solution for the optical depth
$$
\ta(E,z) \cong {A\,\rg\over 2\alpha+3}
\left(z\over\rg\right)^{-2\alpha-3}
\left(E\over 4 m_e c^2\right)^\alpha\,, \eqno(4.6)
$$
valid if $z \gapprox \r0$. In \S~5 we use this result to estimate the
height of the $\gamma$-ray ``photosphere,'' defined as the
value of $z$ such that $\ta=1$. We also calculate
the $\gamma-\gamma$ optical depth for the $\gamma$-ray blazar 3C~279
by focusing on three illustrative scenarios for the propagation
of the $\gamma$-ray: (i) propagation along the disk axis;
(ii) propagation along a line intersecting the disk axis;
and (iii) propagation parallel to the disk axis.

\bigskip
\centerline{\bf 5. APPLICATION TO 3C~279}
\bigskip

The observations of strong fluxes of X-rays and high-energy
$\gamma$-rays from the blazar 3C~279 in June 1991 allows us
to place constraints on the location of the $\gamma$-ray
emission site based on $\gamma-\gamma$ transparency requirements.
Under the assumption that the contemporaneously observed X-rays
were produced by a two-temperature accretion disk, we can use
the expressions developed in \S\S~3 and 4 to calculate the optical
depth traversed by the observed high-energy $\gamma$-rays. Since
variability arguments (Makino et al. 1989) suggest that a significant
fraction of the X-rays were probably produced in a relativistic jet,
the results we will obtain for
the optical depth $\ta$ are upper limits.

The X-ray flux reported by Makino et al. (1993; Fig.~2) during the
outburst had spectral index $\alpha=0.68$ and flux parameter
$F'_0=5.76 \times 10^{-5}\,{\rm s^{-1}\,cm^{-2}}$ (note that
the flux given by their eq.~[1] appears too large by a
factor of $10$). For this value of $\alpha$, we obtain
$y=1.6$ (see eq.~[2.1]) and $\Psi(\alpha)=0.245$ (see Fig.~1),
and for the redshift
of 3C~279 ($\redshift=0.538$), we obtain the luminosity distance
(Lang 1980) $D = H_0^{-1}\,c\,[\redshift + 0.5\,(1-q_0)\,\redshift^2]
=5.65 \times 10^{27}\,$cm assuming $q_0=0.5$ and
$H_0=100\,{\rm km\,s^{-1}\,Mpc^{-1}}$. We can relate $F'_0$
to the disk-frame intensity parameter $I_0$ using
$$
I_0 = {(2-\omega)(1+\redshift)^{3+\alpha}F'_0\over 2\pi}
\left(D\over\rg\right)^2
\left[\left(\r0\over\rg\right)^{2-\omega}
-\left(\rms\over\rg\right)^{2-\omega}\right]^{-1}\,, \eqno(5.1)
$$
according to equations~(2.8) and (2.9).

\medskip
\centerline{5.1. \it Distance to the Gamma-Ray Emission Site}
\medskip

As an application, we consider first the simplest geometrical case,
that of a $\gamma$-ray propagating away from the black hole along
the axis of a two-temperature disk emitting with the power-law X-ray
intensity given by equation~(2.3). This scenario is useful for
determining the minimum height above the disk at which the observed
$\gamma$-radiation could have been produced in 3C~279 in June 1991.
We can calculate the optical depth
to pair production for a $\gamma$-ray of energy $E$ created at height
$z$ above the center of the disk and propagating outward along the
$z$-axis using equation~(4.3). In order to calculate $\ta$, we must
assume values for the inner ($\rms$) and outer ($\r0$) radii of
the two-temperature, X-ray emitting region. In Figure~3 we set
$\rms=6\,\rg$ (accretion onto a Schwarzschild black hole) and
$\r0=30\,\rg$ (a typical value for the outer radius; see SLE and EK),
and plot $\log\ta$ as a function of $\log(M/\msun)$ and $\log(z/\rg)$
for a $\gamma$-ray of energy $E=1\,$GeV using the X-ray data taken
during the June 1991 EGRET flare of 3C~279. The optical depth exceeds
unity in the cross-hatched area, and $\ta=1$ along the heavy solid line
denoting the $\gamma$-ray photosphere, with height $z=\zg$. For reference,
$z = \r0$ along the upper dashed horizontal line, and $z = \rms$ along the
lower dashed line. The slope of the contours depends on how close to the
disk the $\gamma$-rays are produced relative to $\rms$ and $\r0$. For
$z \lapprox \rms$, the contours become increasingly vertical because
the absorption coefficient loses its dependence on $z$ close to the
disk. For $z \gapprox \r0$, the contours make a transition to the
far-field power-law behavior given by equation~(4.6), which reduces to
$\ta\cong 2.4 \times 10^{16}(M/\msun)^{-1}(z/\rg)^{-4.36}$
for the parameters used in Figure~3.

In the case of 3C~279, we find that when $M \gapprox 4 \times 10^{11}\,
\msun$,
the intensity is so low that $\ta < 1$ regardless of the value of
$z$. On the other hand, we see that observable $\gamma$-rays can
be produced quite close to the black hole for reasonable values of
$M$ if $z \gapprox \r0$. For example, if $M=10^9\,\msun$, then
$\ta < 1$ for $z > \zg = 45\,\rg = 2 \times 10^{-3}\,$pc. This suggests
that $\gamma$-rays produced at the top of the evacuated funnel
at the center of a hot accretion disk (perhaps as a consequence
of MHD waves generated in the surrounding hot-disk plasma; see
Becker \& Kafatos 1994), can escape without $\gamma-\gamma$
attenuation by the disk X-rays (although they may suffer attenuation
by scattered X-rays on larger spatial scales; see Dermer \& Schlickeiser
1994).


The precise value of the outer radius $\r0$ of the two-temperature
region is uncertain, but it is likely to fall in the range
$30 \lapprox \r0/\rg \lapprox 100$ (SLE; EK). In Figure~4 we
therefore recompute the $\ta$ contours using the same parameters
as in Figure~3, except $\r0 = 100\,\rg$. Here again $z = \r0$
and $z = \rms$ along the upper and lower dashed lines, respectively.
The differences between Figures~3 and 4 are not dramatic, although
the height of the $\gamma$-ray photosphere has increased from
$\zg=45\,\rg$ to $\zg=70\,\rg$ for $M = 10^9\,\msun$. This is due
to the presence of X-ray photons emitted at large radii, for which
the interaction angle $\theta$ (and therefore the probability of pair
production) is enhanced. In this case, equation~(4.6) for the
approximate far-field
solution reduces to
$\ta\cong 3.6 \times 10^{17}(M/\msun)^{-1}(z/\rg)^{-4.36}$,
in good agreement with Figure~4 for $z \gapprox \r0$.

To investigate the effect of the angular dependence of the
$\gamma-\gamma$ absorption coefficient on the values obtained
for $\ta$, in Figure~5 we plot the height of the $\gamma$-ray
photosphere, $\zg/\rg$, as a function of the black hole mass
$M/\msun$
for $\rms=6\,\rg$ and $\omega=0$ (uniformly-bright disk; {\it dashed
line}) or
$\omega=3$ (two-temperature disk; {\it solid line}). We compare
results obtained
using our two standard values for the outer radius, $\r0=30\,\rg$
({\it open squares}) and $\r0=100\,\rg$
({\it open triangles}). For $\omega=3$, increasing $\r0$ from $30\,\rg$
to $100\,\rg$ causes only a modest increase in the height of the
photosphere as mentioned above. However, going from $\omega=3$
to $\omega=0$ while holding $\r0$ constant results in a substantial
increase in $\zg$ due to the increased amount of X-radiation
impinging on the axially-propagating $\gamma$-rays from large radii,
where the interation angle $\theta$ has its maximum value.
This effect is particularly strong for $\r0=100\,\rg$.

It is also interesting to consider the effect of varying the inner
radius of the disk, $\rms$, since this will also impact on the angular
distribution of the X-rays. In Figure~6 we plot the photosphere height
$\zg/\rg$ as a function of $M/\msun$ for $\rms/\rg=2$, $\r0/\rg=
(30\,,100)$, and
$\omega=(0\,,3)$. These results may represent approximately situations
involving disks around Kerr black holes, for which the radius of
marginal stability can approach $1.2\,\rg$ (Shapiro \& Teukolsky 1983).
In this case equation~(2.2) for the flux $F(R)$ is no longer valid,
although the flux still maintains the approximate power-law radial
dependence $F(R)\propto R^{-3}$ (Eilek 1980). Obviously our model
cannot be rigorously defended in this context since we have completely
neglected relativistic effects, but, bearing this caveat in mind, it is
interesting to note the modest decrease in the height of the photosphere
going from Figure~5 to Figure~6 when $\omega=3$, due to the decrease in
the flux-averaged value of the interaction angle $\theta$. By contrast,
the results depicted in Figures~5 and 6 agree closely when $\omega=0$
(uniformly-bright disk), because then most of the emission comes from
radii $R \gg \rms$, and the value of $\rms$ is therefore irrelevant.

In Table~1 we give values for the height of the $\gamma$-ray photosphere
$\zg/\rg$, along with values for the associated light-crossing timescale
(which may approximate the variability timescale), $\tg\equiv\zg/c$,
for $E=1\,$GeV, $\omega=3$, and various values of the black hole mass
$M/\msun$, the outer disk radius $\r0/\rg$, and the inner disk radius $\rms/\rg$.
Table~1 also includes values for the disk-frame intensity parameter
$I_0$. Using our standard parameters
$M=10^9\,\msun$, $\r0=30\,\rg$, $\rms=6\,\rg$, and $\omega=3$, we find that
$\tg\sim 2\,$d, in agreement with the $\gamma$-ray variability timescale
reported by Kniffen et al. (1993).

\medskip
\centerline{5.2. \it Gamma-Ray Collimation}
\medskip

The dependence of the $\gamma-\gamma$ absorption probability
on the angular distribution of the X-rays suggests that pair production
may help to collimate the high-energy radiation escaping from blazars.
We expect that for given values of the inner and outer disk radii
$\rms$ and $\r0$, the optical depth to pair production for
$\gamma$-rays originating at a fixed location above the center
of the disk will increase with increasing propagation angle $\Phi$,
mainly due to
an increase in the flux-averaged value of the interaction angle
$\theta$, and to a lesser extent because the $\gamma$-rays remain
close to the disk for a longer period of time. We shall continue
to work with X-ray data obtained during the June 1991 flare of
3C~279 in all of the examples.

We can use equation~(4.1) to calculate the optical depth $\ta$ for
$\gamma$-rays of energy $E$ created at a height $z$ above the
center of the disk and propagating at an angle $\Phi$ with respect
to the $z$-axis. In order to explore the angular dependence
of $\gamma-\gamma$ attenuation, in Figures~7 and 8 we plot the
transmission coefficient $\exp(-\ta)$ calculated using the
3C~279 X-ray data as a function of $\Phi$ for $\rms=6\,\rg$,
$E=1\,$GeV, $M=10^9\,\msun$, and various values of $\r0$ and
$z$. The transmission coefficient gives the fraction of the
$\gamma$-rays that would escape to infinity in a given direction
out of those emitted in that direction. We set $\r0=30\,\rg$ in
Figure~7, $\r0=100\,\rg$ in Figure~8, and, to illustrate the
importance of angular effects, we set $\omega=3$ in Figures~7{\it a}
and 8{\it a}, and $\omega=0$ in Figures~7{\it b} and 8{\it b}. Strong
``focusing'' of the escaping $\gamma$-rays along the $z$-axis is
apparent in all cases, and, for $\omega=3$, the half-width of the
angular distribution, $\Delta\Phi$, falls in the range
$10^\circ \lapprox \Delta\Phi \lapprox 20^\circ$. Similar results
are obtained for $\Delta\Phi$ when $\omega=0$ (Figs.~7{\it b}
and 8{\it b}), but the overall level of attenuation is much
greater in these cases due to the enhancement of the X-ray
emission produced at large radii in the disk. In Table~1 we
give values for $\Delta\Phi$ evaluated at the photosphere, $z=\zg$,
for various model parameters. Our results for $\Delta\Phi$
are comparable to the observationally determined values (e.g., Padovani
\& Urry 1992), and we therefore conclude that for acceptable disk
parameters, $\gamma-\gamma$ attenuation alone is enough to produce
the degree of $\gamma$-ray collimation implied by observations which
associate extragalactic $\gamma$-ray emission with blazars.

\medskip
\centerline{5.3. \it Shape of the Gamma-Ray Photosphere}
\medskip

The high-energy $\gamma$-rays observed from many blazars are
emitted nearly parallel to the axis of the jet, as is evidenced
particularly by the common occurrence of apparent superluminal motion
in the sources. It is therefore interesting to calculate,
within the context of our model, the $\gamma-\gamma$ optical
depth for a $\gamma$-ray created at radius $R$ and height $z$,
and subsequently propagating parallel to the $z$-axis of the disk,
which can be obtained by setting $\Phi=0$ in equation~(4.1). It is
particularly interesting to determine the shape of the $\gamma$-ray
photosphere by varying the value of the $\gamma$-ray creation radius
$R$.

In Figure~9, we plot the photosphere height $\zg/\rg$
calculated using the 3C~279 X-ray data as a function of $R/\rg$
for various values of $M/\msun$, assuming our standard parameters
$\r0=30\,\rg$, $\rms=6\,\rg$, and $\omega=3$. We find that the
height of the photosphere increases
dramatically as $M$ decreases, and the shape of the photosphere
becomes quite concave. Conversely, for large values of $M$,
the photosphere becomes nearly flat. We interpret this behavior
as follows. For large values of $M$, the disk-frame intensity parameter
$I_0$ is small, and therefore the photosphere is located relatively
close to the disk, with $\zg/\r0 \lapprox 1$. In this case, most of
the $\gamma-\gamma$ absorption takes place so close to the disk that
the flux-averaged value of the interaction angle $\theta$ is nearly
$90^\circ$ at all points in the photosphere, and therefore the height
of the photosphere varies little across the disk. On the other hand,
as $M$ decreases, $I_0$ becomes large, and consequently the photosphere
lies further from the disk, with $\zg/\r0 \gapprox 1$. In this case,
the flux-averaged value of $\theta$ increases significantly going
from the center of the photosphere to the edge, and therefore the
height of the photosphere must increase rapidly in order to keep
$\ta=1$, as seen in Figure~9.

\bigskip
\centerline{\bf 6. DISCUSSION}
\bigskip

Although the physics of particle creation and acceleration far from
the turbulent plasma accreting onto a black hole is poorly understood,
there are a number of relatively well-understood plasma and accretion
instabilities that can give rise to efficient particle acceleration within
the disk itself, or within the evacuated funnel region at the center of
the disk. For example, resonant turbulent acceleration via interactions
between particles and (linear or nonlinear) Alfv\'en waves can in principle
lead to the production of electrons with very high Lorentz factors, and to
the associated emission of high-energy inverse-Compton $\gamma$-rays (e.g.,
Becker \& Kafatos 1994; Becker, Kafatos, \& Maisack 1994). However, in order
for any of these mechanisms to create {\it observable} $\gamma$-rays, the
optical depth to pair production on the background UV and X-ray photons
must be small enough for the $\gamma$-rays to escape from the source.

In this paper, we have attempted to bridge the gap between theory
and observation by developing useful constraints based on the simple
hypothesis that the power-law X-ray spectrum originates in the inner
region of a two-temperature accretion disk. Such disks commonly provide
good fits to the hard X-ray spectra of AGNs. In \S~3 we developed a
general expression for the $\gamma-\gamma$ absorption coefficient $\ab$
for $\gamma$-rays propagating above a two-temperature disk in a plane
that includes the symmetry axis, and in \S~4 we derived an
approximate far-field solution for the $\gamma-\gamma$ optical
depth $\ta$ for
$\gamma$-rays created above the center of the disk and propagating
outward along the disk axis, taking full account of the angular
dependence of the X-ray distribution. In \S~5 we applied our results
to 3C~279 using the X-ray data reported by Makino et al. (1993) during the June
1991 $\gamma$-ray flare, and demonstrated that the observed
$\gamma$-rays could have been produced within $\lapprox 45\,GM/c^2$
of a black hole of mass $M \gapprox 10^9\,\msun$. This is close enough
to the central source to establish a direct connection between the
production of the $\gamma$-rays and the accretion of material onto
the black hole. In particular, our results suggest that the observed
$\gamma$-rays may have been produced near the top of the evacuated funnel
at the center of a hot, two-temperature accretion disk. In this scenario,
the relativistic electrons that Compton-upscatter UV photons to produce
the $\gamma$-rays are themselves produced as a consequence of
photon-photon collisions in the funnel. The absence of protons
in the funnel due to the centrifugal barrier allows the presence
of Alfv\'en waves with wavelengths short enough to effectively
accelerate the electrons (Becker, Kafatos, \& Maisack 1994).

In order to explore the observational implications of the angular
dependence of $\gamma-\gamma$ pair production, in \S~5 we also
computed $\ta$ for $\gamma$-rays born along the disk axis but
propagating at an angle $\Phi$ with respect to it. The results
for the half-angle ($\Delta\Phi\lapprox 15^\circ$) indicated in
Figures~7 and 8 and in Table~1 suggest that pair production in
the X-ray field of the accretion disk may be strong enough to
explain the observed degree of $\gamma$-ray alignment in blazars.
Furthermore, the variability (light-crossing) timescale
$\tg\equiv\zg/c$ we derive for black hole mass $M \sim 10^9\,\msun$ is in
good agreement with the 2-day $\gamma$-ray variability of 3C~279
reported by Kniffen et al. (1993).

The results presented in \S\S~4 and 5 suggest that $\gamma-\gamma$
pair production can help to ``focus'' an intrinsically isotropic
source of $\gamma$-rays so that the high-energy radiation escapes
preferentially along the disk axis. Whether or not the $\gamma$-rays
actually are produced isotropically is a separate question. If they
are, then several additional considerations come into play, regarding
(i) the self-opacity of the $\gamma$-rays;
(ii) the fate of the $\gamma$-ray energy emitted off-axis; and
(iii) the required total luminosity.

\medskip
\centerline{ 6.1. \it $\gamma-\gamma$ Self-Opacity}
\medskip

Whether or not the high-energy $\gamma$-rays are produced by an
intrinsically isotropic source, their self-interaction optical
depth $\tau_{\rm self}$ cannot exceed unity if they are
to be observed. The most severe self-interaction constraint is
obtained in the case of isotropic emission. Assuming a monochromatic,
isotropic distribution of $\gamma$-rays with energy
$E_0 > m_e c^2$ and intensity $\in = B\,\delta(E-E_0)$, integration
of equation~(3.3) yields for the self-interaction absorption coefficient
$$
\alpha_{\rm self} = {8\,\pi\,B \over c\,E}
\left(E \over m_e c^2\right)^{-4}
\int_0^{\beta_{\rm max}} \si(\beta)\,{2\,\beta\,d\beta
\over (1-\beta^2)^3}\,, \eqno(6.1)
$$
where $\si(\beta)$ is given by equation~(3.1) and
$$
\beta_{\rm max} = \left[1-\left(E\over m_e c^2\right)^{-2}\right]^{1/2}\,.
\eqno(6.2)
$$
If the $\gamma$-ray source is an isotropically emitting sphere of
radius $R_*$, then the intensity parameter $B$ is related to the
$\gamma$-ray luminosity $L_\gamma$ via
$$
B={L_\gamma \over (2 \pi R_*)^2}\,, \eqno(6.3)
$$
and we obtain for the self-interaction $\gamma-\gamma$ optical depth
$$
\tau_{\rm self} \equiv R_*\,\alpha_{\rm self}
={2\,L_\gamma\,\sig \over \pi R_* c E}
\left\{\left(E \over m_e c^2\right)^{-4}
\int_0^{\beta_{\rm max}} {\si(\beta) \over \sig}
{2\,\beta\,d\beta \over (1-\beta^2)^3}\right\}\,.
\eqno(6.4)
$$
An isotropic $\gamma$-ray source can be ruled out if
$\tau_{\rm self}$ exceeds unity for reasonable values of $R_*$. The
first factor on the right-hand side of equation~(6.4) is essentially
the standard $\gamma$-ray compactness, and the quantity in
braces becomes $\sim 10^{-6}$ for $E=1\,$GeV. Taking $L_\gamma=10^{48}\,
{\rm ergs\,s}^{-1}$ for 3C~279 and setting $\tau_{\rm self} < 1$ yields
the transparency requirement $R_* \gapprox 10^{10}\,$cm, and therefore
self-interaction of the GeV radiation in 3C~279 is not likely to be
important. A similar calculation by McNaron-Brown et al. (1995) indicates
that the self-interaction of radiation in the energy range $0.05-10\,$MeV
(observed by OSSE) is also negligible.

\medskip
\centerline{ 6.2. \it Fate of the Off-Axis Radiation}
\medskip

If the $\gamma$-rays are produced isotropically, then only those
emitted close to the disk axis will escape from the X-ray field
of the accretion disk. The radiation emitted further from the axis
will be degraded in a pair-Compton cascade (Protheroe, Mastichiadis,
\& Dermer 1992), with most of the luminosity residing ultimately in
a broad MeV feature. In this view, objects such as 3C~279 (while
in $\gamma$-ray flare mode) will appear as bright sources of
MeV emission when viewed along lines of sight
$\gapprox 20^\circ$ from the disk axis. Centaurus~A may be an example
of such an object (Skibo, Dermer, \& Kinzer 1995).

The luminosity required to power isotropic $\gamma$-ray emission,
$L_\gamma=10^{48}\,{\rm ergs\,s}^{-1}$, is orders of magnitude
higher than that required if the emission is beamed. This is
perhaps the simplest argument against isotropic emission, but
it is especially compelling when coupled with the strong
circumstantial evidence in favor of relativistic bulk motion.
However, there is apparently no strong {\it a priori} reason to rule out isotropic emission, since
attenuation due to collisions with X-rays emitted by the
disk is sufficient to produce the observed degree of alignment.
In any event, our
results regarding $\gamma-\gamma$ collimation do not rely on the existence
of an isotropic $\gamma$-ray source. The mechanism we consider here can work
in concert with other sources of collimation, such as relativistic bulk
motion or geometrical focusing.

\medskip
\centerline{ 6.3. \it Attenuation by Scattered X-Rays}
\medskip

The backscattering of X-rays produced in the accretion disk by
electrons located above the disk tends to isotropize the X-ray
distribution, and this can have a strong effect on the results
obtained for the $\gamma-\gamma$ absorption coefficient. We
believe that our neglect of a scattered component to the X-ray
distribution is reasonable due to the proximity of the $\gamma$-ray
production site to the black hole in our model. It has been shown
(Dermer \& Schlickeiser 1994) that accretion disk photons will
dominate over scattered X-rays within a distance of $\sim 0.01-0.1\,$pc
from the central source, which is much larger than the typical height
of the $\gamma$-ray photosphere, $\zg$, obtained here (see
Table~1). However, the problem of $\gamma-\gamma$ attenuation
by scattered X-rays on much larger spatial scales is a serious issue
that must be addressed using detailed calculations that determine
self-consistently the angle and energy distributions of both the
scattered radiation and the electron-positron pairs at large distances
from the black hole. Such a calculation is beyond the scope of this
paper, but it suggests a promising direction for future work.

Our results indicate that if the emission height $z$ is held fixed,
then the optical depth increases with increasing $\gamma$-ray energy
$E$ as $\ta\propto E^\alpha$, and therefore the height of the
$\gamma$-ray photosphere $\zg$ increases with increasing $E$.
If we arbitrarily set the emission height for all of the $\gamma$-rays
equal to the value of $\zg$ obtained for $E=1\,$GeV, then the observed
spectrum will be attenuated by
a factor $\sim 7$ going from $E=1\,$GeV to $E=5\,$GeV. It is
unclear whether such a steepening actually exists in the spectrum
of 3C~279 (Hartman et al. 1992). Finally, for energies
$E\gapprox 5\,$GeV, the interaction with UV radiation produced
in the cool outer region of the disk would also become important,
further raising the height of the photosphere.

To summarize our main points, application of our general expressions
describing $\gamma-\gamma$ attenuation above X-ray emitting accretion
disks leads us to conclude that in the case of 3C~279,
(i) $\gamma$-ray transparency (to disk-generated X-rays) is achieved
if the $\gamma$-rays are produced at height $z \gapprox \r0$;
(ii) $\gamma-\gamma$ attenuation is strong enough to explain the
observed association of high-energy $\gamma$-ray emission from
AGNs exclusively
with blazars, and
(iii) the observed $\gamma$-rays may be produced
in active plasma located above the funnel at the center
of a two-temperature accretion disk.

We gratefully acknowledge the remarks provided by the anonymous
referee, whose insightful comments contributed significantly to the paper. This work was
partially supported by a NASA Cycle~3 CGRO Guest Investigator grant.

\vfil
\eject

\centerline{\bf REFERENCES}
{\refs

Becker, P. A., \& Kafatos, M. 1994, in The Second {\it Compton}
Symp., ed. C. E. Fichtel, N. Gehrels, \& J. P. Norris,
(New York: AIP), 620

Becker, P. A., Kafatos, M., \& Maisack, M. 1994, ApJS, 90, 949

Bednarek, W. 1993, A\&A, 278, 307

Begelman, M. C., Blandford, R. D., \& Rees, M. J. 1984,
Reviews of Modern Physics, 56, 255

Blandford, R. D. 1993, in Proc. {\it Compton Gamma-Ray Observatory}
Symp., ed. M. Friedlander, N. Gehrels, \& D. J. Macomb (New York:
AIP), 533

Dermer, C. D., \& Schlickeiser, R. 1993, ApJ, 416, 458

Dermer, C. D., \& Schlickeiser, R. 1994, ApJS, 90, 945

Eilek, J. A. 1980, ApJ, 236, 664

Eilek, J. A., \& Kafatos, M. 1983, ApJ, 271, 804 (EK)

Fichtel, C. E., et al. 1994, The First Energetic Gamma-Ray
Experiment Telescope (EGRET) Source Catalog, (Washington, DC: NASA)

Gould, R. J., \& Schr\'eder, G. P. 1967, Phys. Rev., 155, 1404

Hartman, R. C., et al. 1992, ApJ, 385, L1

Hermsen, W., et al. 1993, A\&AS, 97, 97

Jauch, J. M., \& Rohrlich, F. 1955, Theory of Photons
and Electrons (Reading: Addison-Wesley)

Kniffen, D. A., et al. 1993, ApJ, 411, 133

Lang, K. R. 1980, Astrophysical Formulae (New York: Springer-Verlag)

Makino, F., et al. 1989, ApJ, 347, L9

Makino, F., et al. 1993, in Frontiers of Neutrino Astrophysics, ed.
Y. Suzuki \& K. Nakamura (Tokyo: Universal Press), 425

McNaron-Brown, K., et al. 1994, in The Second {\it Compton}
Symp., ed. C. E. Fichtel, N. Gehrels, \& J. P. Norris,
(New York: AIP), 587

McNaron-Brown, K., et al. 1995, ApJ, submitted

Padovani, P., \& Urry, C. M. 1992, ApJ, 387, 449

Pringle, J. E. 1981, ARA\&A, 19, 137

Protheroe, R. J., Mastichiadis, A., \& Dermer, C. D. 1992,
Astroparticle Phys., 1, 113

Shakura, N. I., \& Sunyaev, R. A. 1973, A\&A, 24, 337

Shapiro, S. L., Lightman, A. P., \& Eardley, D. M. 1976, ApJ,
204, 187 (SLE)

Shapiro, S. L., \& Teukolsky, S. A. 1983, Black Holes, White Dwarfs,
and Neutron Stars (New York: John Wiley)

Skibo, J. G., Dermer, C. D., \& Kinzer, R. L. 1995, ApJ, in press

von Montigny, C., et al. 1995, ApJ, 440, 525

\vfil
\eject
\centerline{\bf FIGURE CAPTIONS}

\noindent

Fig.~1. -- The cross section integral function $\Psi(\alpha)$
(eq.~[3.7]). For the 3C~279 data, $\alpha=0.68$ (see Makino
et al. 1993), and $\Psi=0.245$.

Fig.~2. -- Schematic representation of the $\gamma$-ray propagation
geometry. The $\gamma$-ray is located at distance $\rho$ from the
origin and polar angle $\delta$ and propagates in a plane that includes
the $z$-axis. Attenuation of the $\gamma$-rays occurs as a result of
collisions with X-rays generated at all points on the disk. The interaction
angle between the photons is $\theta$, and the angle between the
$\gamma$-ray trajectory and the $z$-axis is $\Phi$.

Fig.~3. -- Contour plot of the log of the $\gamma-\gamma$ optical
depth $\ta$ (eq.~[4.3]) for $\gamma$-rays with energy $E=1\,$GeV
created at height $z$ above the center of a two-temperature disk
($\omega=3$) with outer radius $\r0=30\,\rg$ and inner radius
$\rms=6\,\rg$. The upper and lower dashed horizontal lines denote
$z=\r0$ and $z=\rms$, respectively, and the $\gamma$-rays propagate
outward along the $z$-axis. The curves were computed using the X-ray
data for 3C~279 taken during the June 1991 EGRET flare by Makino et
al. (1993).

Fig.~4. -- Same as Fig.~3, except $\r0 = 100\,\rg$.

Fig.~5. -- Height of the $\gamma$-ray photosphere, $\zg/\rg$,
plotted as a function of the black hole mass $M/\msun$ for
$\rms=6\,\rg$ and $\omega=0$ ({\it dashed line}), $\omega=3$
({\it solid line}), $\r0=30\,\rg$ ({\it open squares}),
$\r0=100\,\rg$ ({\it open triangles}). The curves were
computed using the June 1991 3C~279 X-ray data.

Fig.~6. -- Same as Fig.~5, except $\rms = 2\,\rg$.

Fig.~7. -- Transmission coefficient $\exp(-\ta)$ plotted
as a function of the propagation angle $\Phi$ using the
June 1991 3C~279 X-ray data for $M=10^9\,\msun$,
$\r0=30\,\rg$, $\rms=6\,\rg$ and ({\it a}) $\omega=3$,
({\it b}) $\omega=0$. The $\gamma$-rays are created at the
height $z/\rg$ indicated for each curve.

Fig.~8. -- Same as Fig.~7, except $\r0 = 100\,\rg$.

Fig.~9. -- Height of the $\gamma$-ray photosphere $\zg/\rg$
plotted as a function of the creation radius $R/\rg$ for
$\gamma$-rays propagating parallel to the $z$-axis using
the June 1991 3C~279 X-ray data, assuming $\r0=30\,\rg$,
$\rms=6\,\rg$, and $\omega=3$. The value of the black hole
mass $M/\msun$ is indicated for each curve.

\bye